\documentstyle[psfig,conf_iap,lscape]{article}
\begin{document}
%

\def \vc #1{{\mbox{\boldmath $#1$}}}
\def\rmk{{\rm k}}

\def\pg{{\vc p}}
\def\eg{{\vc e}}
\def\boldk{{ k}}
\def\boldx{{ x}}
\def\boldg{{ g}}
\def\nablag{{\vc \nabla}}
\def\thetag{{\vc \theta}}
\def\betag{{\vc \beta}}
\def\alphag{{\vc \alpha}}
\def\deltag{{\vc \delta}}
\def\gammag{{\vc \gamma}}
\def\epsilong{{\vc \epsilon}}
\def\taug{{\vc \tau}}

\def\Journal#1#2#3#4{{#1} {\bf #2}, #3 (#4)}

\def\NCA{\em Nuovo Cimento}
\def\NIM{\em Nucl. Instrum. Methods}
\def\NIMA{{\em Nucl. Instrum. Methods} A}
\def\NPB{{\em Nucl. Phys.} B}
\def\PLB{{\em Phys. Lett.}  B}
\def\PRL{\em Phys. Rev. Lett.}
\def\PR{\em Phys. Rev.}
\def\pr{\em Phys. Rep.}
\def\PRD{{\em Phys. Rev.} D}
\def\ZPC{{\em Z. Phys.} C}
\def\apj{ApJ}
\def\nat{\em Nature}
\def\aa{A\& A}
\def\mnras{MNRAS}
\def\proclon{Proc. Roy. Soc. London}

\heading{%
%
Cosmic shear review\\
%
} 
\par\medskip\noindent
\author{%
Ludovic Van Waerbeke$^{1}$, Ismael Tereno$^{1,2}$, Yannick Mellier$^{1,3}$, Francis Bernardeau$^{4}$
}
\address{%
Institut d'Astrophysique de Paris, CNRS, 98bis Boulevard Arago, 
       F--75014 Paris.
}
\address{%
University of Lisboa, Dept. of Physics, 1749-016, Lisboa.
}
\address{%
Obs. de Paris/LERMA, 77 Av. Denfert-Rochereau, F--75014 Paris.
}
\address{%
Service de Physique Th\'eorique de Saclay,
       F--91191 Gif sur Yvette Cedex.
}

\begin{abstract}
We present an overview of all the cosmic shear results obtained so far.
We focus on the 2-point statistics only. Evidences supporting the
cosmological origin of the measured signal are reviewed, and issues related
to various systematics are discussed.
\end{abstract}
\section{Introduction}
The cosmic shear is a gravitational lensing effect caused by the
large scale structures in the universe on the distant galaxies. Its net effect is
to distort coherently the galaxy images over large angular scales and to change the
focusing properties of the light beam (galaxies appear larger or smaller depending
on the intervening mass density). Only the former effect is 'easily' measurable,
and is the one we will discuss here. The latter, which has been measured in some
cluster lensing cases, is much more challenging for cosmic shear
purposes.

The measurement of the amplitude of
the cosmic shear as a function of scale is a direct indicator of the projected
mass power spectrum, convolved with a selection function which only depends on the
cosmological parameters and the redshift distribution of the sources. It is
therefore a tool for evaluating the mass distribution in the universe, as well
as for measuring the cosmological parameters. In the following, we first outline the
theory of cosmic shear, and its link to measurable quantities in Section2.
We discuss the measurements and the systematics in Section 3.
Section 4 is a discussion of the cosmological parameters
measurements and the still open problems associated with it.
\section{Theory}
In the presence of mass inhomogeneities, a light ray is deflected, and it is
observed at an angular position $\betag$ on the sky instead of at its intrinsic
location $\thetag$. The mapping between the two position angles defines
the amplification matrix $\cal A$

\begin{equation}
{\cal A}={\partial \beta_i\over \partial \theta_j}=\left(\matrix{1-\kappa-\gamma_1
\ \ \ \ \ \ \ \gamma_2 \cr \gamma_2 \ \ \ \ \ \ \  1-\kappa+\gamma_1}\right),
\end{equation}
where the convergence $\kappa$ and the shear $\gammag=(\gamma_1,\gamma_2)$
are given by the second derivatives of the projected gravitational
potential $\varphi$:

\begin{equation}
\kappa={1\over 2}\left(\varphi_{,11}+\varphi_{,22}\right)\ ; \ \
\gamma_1={1\over 2}\left(\varphi_{,11}-\varphi_{,22}\right)\ ; \ \
\gamma_2=\varphi_{,12}.
\end{equation}

\begin{figure}
\centerline{\vbox{
\psfig{figure=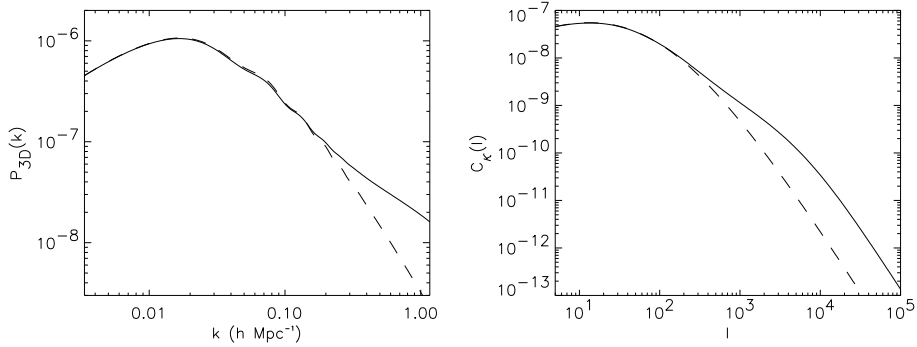,height=5.cm}
}}
\caption[]{The left panel is a 3-dimensional mass power spectrum for the
linear (dashed) and non-linear (solid) regimes when baryons are included.
A value of $\Omega_b=0.05$
was used. The right panel shows the induced convergence power spectrum
(Eq.\ref{pofkappa}) for the two dynamical regimes. Other parameters
are $\Omega_{\rm cdm}=0.25, \Omega_\Lambda=0.7, \sigma_8=0.9, h=0.7,
z_{\rm source}=0.8$.
\label{pdek.ps}}
\end{figure}

The central quantity in cosmic shear analysis is the convergence power spectrum
$P_\kappa(k)$, which relates any cosmic shear two point statistics
to the cosmological parameters and the 3-dimensional mass power
spectrum $P_{3D}(k)$:

\begin{equation}
P_\kappa(k)={9\over 4}\Omega_0^2\int_0^{w_H} {{\rm d}w \over a^2(w)}
P_{3D}\left({k\over f_K(w)}; w\right)
\left[ \int_w^{w_H}{\rm d} w' n(w') {f_K(w'-w)\over f_K(w')}\right]^2,
\label{pofkappa}
\end{equation}
where $f_K(w)$ is the comoving angular diameter distance out to a
distance $w$ ($w_H$ is the horizon distance), and $n(w(z))$ is the
redshift distribution of the sources. The mass power spectrum
$P_{3D}(k)$ is evaluated in the non-linear regime
\cite{smith02}, and $k$ is the 2-dimensional wave vector
perpendicular to the line-of-sight.
Figure \ref{pdek.ps} is an example of 3-dimensional and convergence power
spectra for comparison. A fair amount of baryons was included (using CAMB
\cite{lewis}), in order to show that the baryon oscillations, which
are clearly visible on the 3D spectrum, are severely diluted in the
projected spectrum.

\begin{figure}
\centerline{\vbox{
\psfig{figure=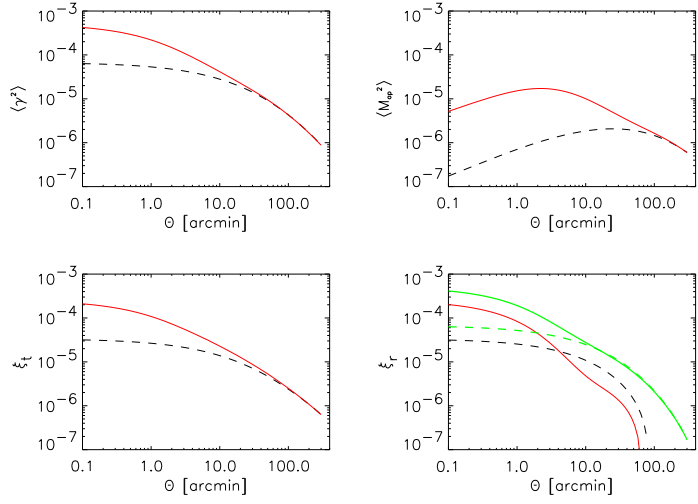,height=7.cm}
}}
\caption[]{Lensing statistics predictions for the cosmological model used in
Figure \ref{pdek.ps}. Both linear (dashed) and non-linear (solid lines) regimes
are represented. On the bottom-right plot, the thick dashed and solid lines
are the full shear correlation function.
\label{model.ps}}
\end{figure}
The cosmic shear effect is measured from the ellipticity $\eg$ of the
galaxies, which is assumed to be fully described by their second
order moments ${\cal M}$ of the surface brightness $I(\thetag)$:

\begin{equation}
{\cal M}_{ij}=
\int I\left( \thetag \right) \theta_i \ \theta_j
\ {\rm d}^2\theta  \ ; \ \ \
\eg=\left({{\cal M}_{11}-{\cal M}_{22}\over {\rm Tr}({\cal M})} , \ \ 
{2{\cal M}_{12}\over {\rm Tr}({\cal M})}\right).
\end{equation}
To first order in the lensing amplitude, $\eg$ is an unbiased estimate of the
shear $\gammag$. The three most commonly measured 2-point statistics are
respectively the
shear top-hat variance \cite{me91,b91,k92}, the aperture mass
variance \cite{k94,sch98} and the shear correlation
function \cite{me91,b91,k92}. They are all different
measurements of the same quantity, the convergence power spectrum
$P_\kappa(k)$:
\begin{equation}
\langle\gamma^2\rangle={2\over \pi\theta_c^2} \int_0^\infty~{{\rm d}k\over k} P_\kappa(k)
[J_1(k\theta_c)]^2 \ ; \ \ 
\langle M_{\rm ap}^2\rangle={288\over \pi\theta_c^4} \int_0^\infty~{{\rm d}k\over k^3}
P_\kappa(k) [J_4(k\theta_c)]^2, \nonumber
\label{theomap}
\end{equation}
\begin{eqnarray}
\xi(\theta_c)&=&\langle\gammag(r)\cdot\gammag(r+\theta_c)\rangle_r={1\over 2\pi} \int_0^\infty~{\rm d} k~
k P_\kappa(k) J_0(k\theta_c) \cr 
\left(\matrix{\xi_t(\theta_c)\cr \xi_r(\theta_c)}\right)&=&{1\over 4\pi} \int_0^\infty~{\rm d} k~
k P_\kappa(k) \left[J_0(k\theta_c)\pm J_4(k\theta_c)\right] 
\label{theogg}
\end{eqnarray}
where $J_n$ is the Bessel function of the first kind, and $\theta_c$ is
the smoothing radius (or the pair separation for the shear correlation
functions). 
\begin{figure}[t]
\centerline{\vbox{
\psfig{figure=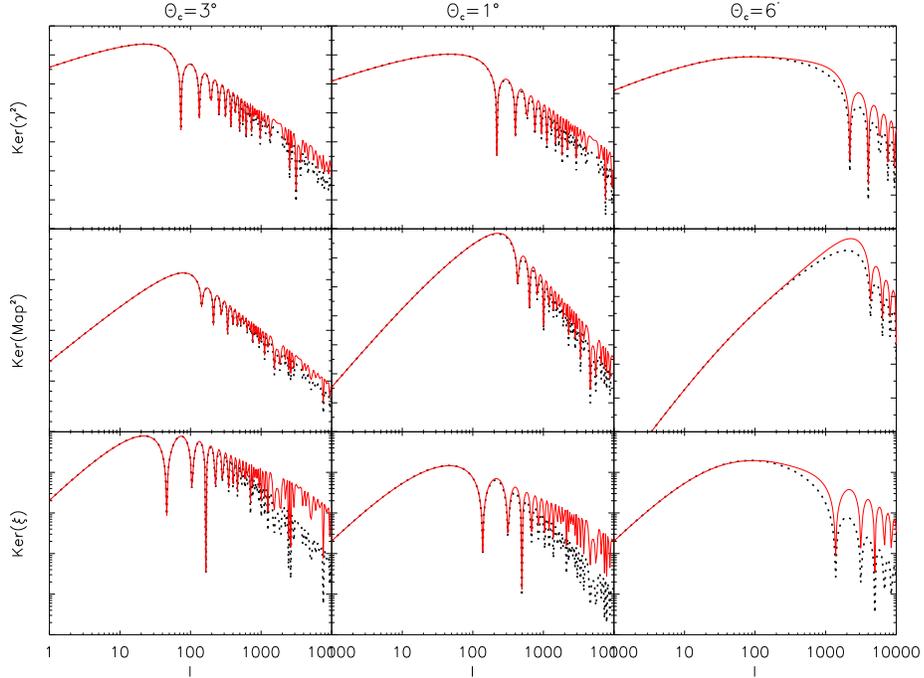,height=9.cm}
}}
\caption[]{Integrants of the different lensing statistics (Eqs.(\ref{theomap},\ref{theogg}))
for the linear (dotted line) and non-linear (solid line) regimes.
From top to bottom: top-hat variance, aperture mass and total correlation function.
The vertical axis is arbitrary, and the three columns correspond to $\theta_c=3 {\rm deg}$,
$1 {\rm deg}$ and $6'$ respectively.
\label{kernel.ps}}
\end{figure}
Figure \ref{model.ps} is an example of lensing quantities predictions for the
same cosmology as in Figure \ref{pdek.ps}. It shows the importance of the non-linear
corrections for scales below $30'$ to $1$ degree. An illustration of the difference
between the linear and non-linear regimes is done by plotting the integrants
of Eqs.(\ref{theomap}, \ref{theogg}) as in Figure \ref{kernel.ps}. Figures
\ref{model.ps} and \ref{kernel.ps} make it clear that the lensing signal is
dominated by the first few peaks in the smoothing kernel, with a transition
linear/non-linear around a smoothing scale of $1$ degree, depending slightly on
the statistic under interest.

\section{Observations}

\subsection{Two-point statistics}
\begin{figure}[t]
\centerline{\vbox{
\psfig{figure=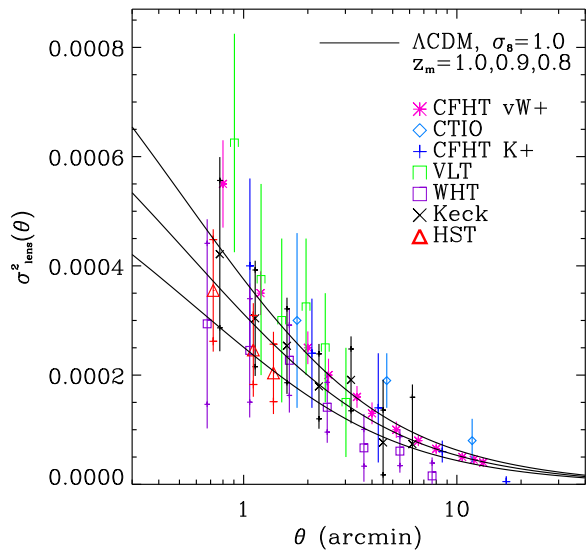,height=8.cm}
}}
\caption[]{Compilation of recent results of top-hat shear variance
measurements from several groups \cite{ref02}.
\label{ref02.eps}}
\end{figure}
\begin{figure}
\centerline{
\psfig{figure=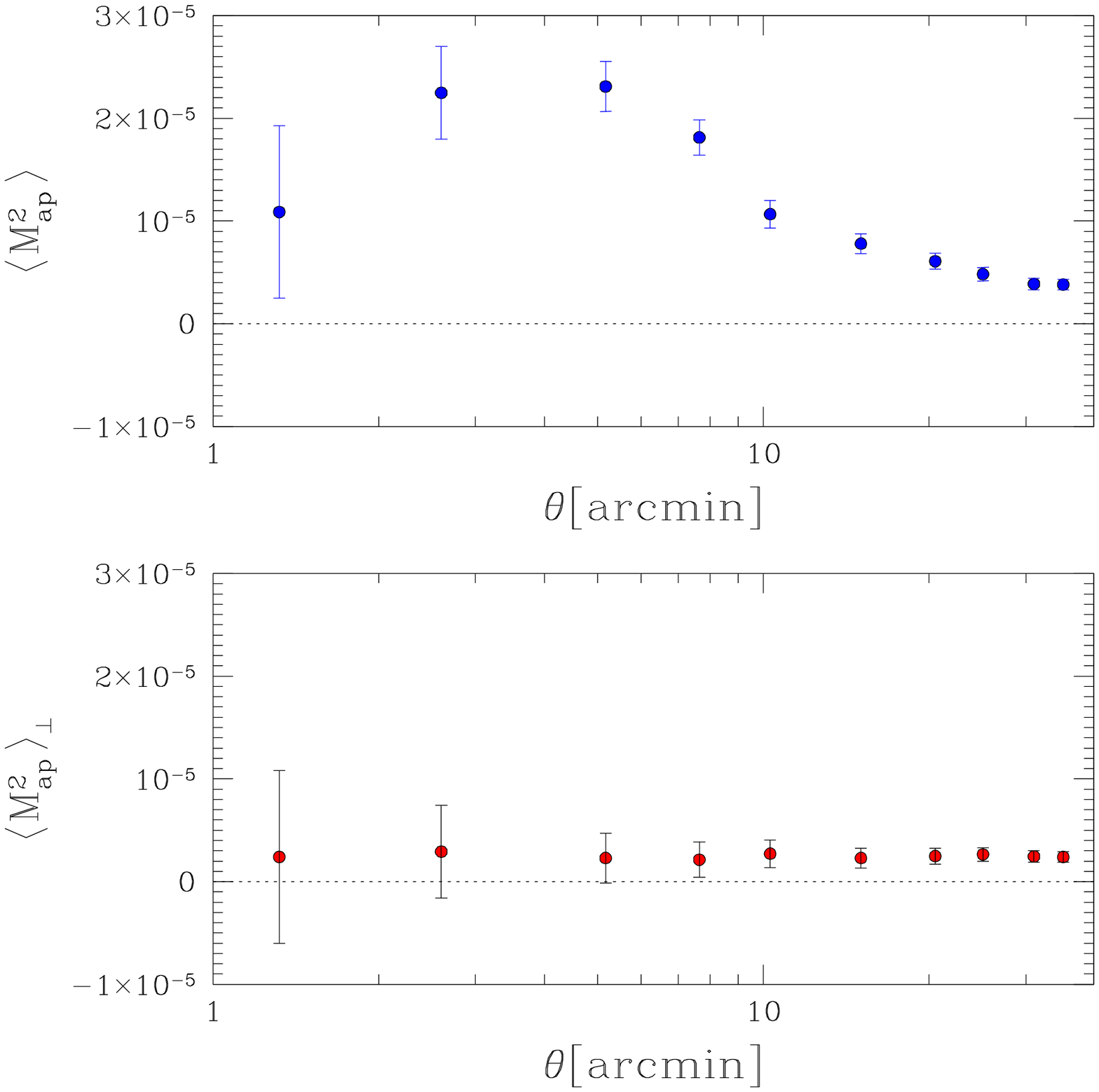,height=7cm}
\psfig{figure=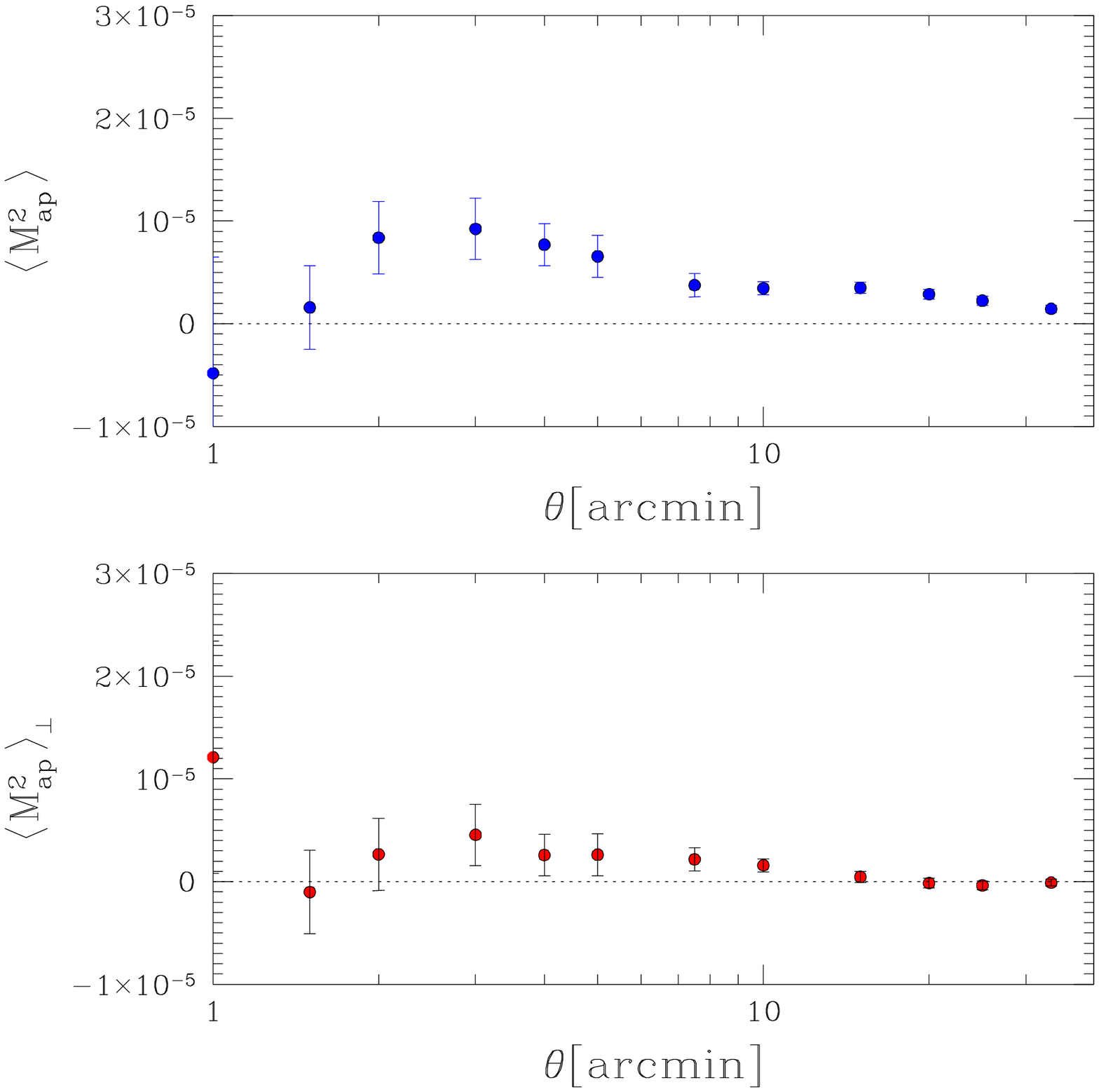,height=7cm}}
\caption{Left: $E$ (top) and $B$ (bottom) modes measured in the VIRMOS
survey. Right: $E$ (top) and $B$ (bottom) modes measured in the
RCS survey. The $B$ mode is low and the $E$ mode compatible
with the predictions for the aperture mass statistics.
\label{mapplot}}
\end{figure}

There are
now several evidences of the cosmological origin of the measured signal:

(a) The consistency of the shear excess variance measured from different telescopes,
at different depths and with different filters. This is summarized on Figure
\ref{ref02.eps}.

(b) On a single survey, the self consistency of the different types of lensing
statistics as given in Eqs.(\ref{theomap},\ref{theogg}) \cite{vwal01}.

(c) The $E$, $B$ modes decomposition separates the lensing signal into
curl and curl-free modes \cite{crit02}. It is expected, and it can also
be quantified on the star field \cite{vwal02}, that residual systematics equally
contribute to $E$ and $B$, while the lensing signal should be present ONLY
in the $E$ mode. This is a consequence that gravity derives from a true
scalar field \cite{steb96}.
The $E$ mode is identical to the aperture mass statistic Eq.(\ref{theomap})
\cite{k94,sch98}, while the $B$ mode can be computed in the same way by  
rotating each galaxy by $45$ degrees. The $E$ and $B$ modes have been measured in
several surveys
\cite{vwal01,vwal02,pen02,hoekstra02,brown02}, and support the cosmological origin of the
signal, showing also the already small amount of residual systematics
achieved with today's technology. Figure \ref{mapplot} shows such measurements
for the VIRMOS--DESCART \footnote{http://www.astrsp-mrs.fr}
\footnote{http://terapix.iap.fr/DESCART} and RCS
\footnote{http://www.astro.utoronto.ca/~gladders/RCS/} surveys.

(d) The lensing signal is expected to decrease for low redshift sources,
as consequence of the lower efficiency of the the gravitational
distortion. This decrease
of the signal has been observed recently for the first time, when comparing
the VIRMOS survey aperture mass \cite{vwal02} which has a source mean redshift
around $0.9$ to the RCS which has a source mean redshift around $0.6$. The
expected decrease in signal amplitude is about $2$, which is what is observed
(see Figure \ref{mapplot}).

(e) Space images provide in principle a systematics-free environment, and even if
the observed areas are still smaller than ground based observations, space
data provide ideal calibrations of the cosmic shear signal
\cite{rhodes01,hammerle01,ref02},
which are in excellent agreement with ground based measurements.

\subsection{Systematics}

\begin{figure}
\centerline{\vbox{
\psfig{figure=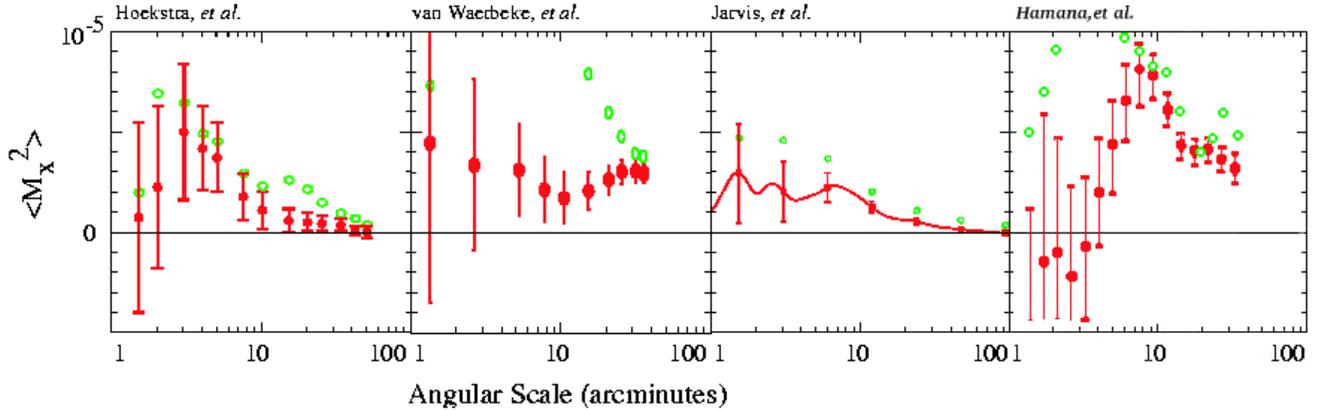,height=6.cm}
}}
\caption[]{Plot showing the relative amplitude of the $E$ and $B$ modes
(points without and with error bars respectively) for the surveys where
the aperture mass has been measured \cite{hoekstra02,vwal02,jar02,ham02}
(picture taken from \cite{jar02}, and extended) The result of
\cite{hoekstra02} is for the full magnitude range, while in Figure
\ref{mapplot}, right panel, it is for the galaxies used for the cosmic shear analysis.
\label{bmode4ex.ps}}
\end{figure}
As we said, any 2-points statistic can be decomposed
into the so-called $E$ and $B$
modes channels, which separate the cosmological signal from the systematics
\cite{crit02,pen02}. Figure \ref{bmode4ex.ps} shows
\footnote{The $B$ mode peak
at $10'$ in \cite{ham02} is due to a PSF correction error over the mosaic.
It is gone when the proper correction is applied, Hamana, {\it private communication}.}
the $E$ and $B$ modes
that have been measured so far, using the aperture mass only (this is
the statistic which provides an unambiguous $E$ and $B$ separation
\cite{pen02}). The two deepest
surveys have large scale
$B$ mode contamination \cite{vwal02,ham02}, and the two shallow surveys have
small scale contamination \cite{hoekstra02,jar02}.

The source of systematics is still unclear. It could come in part from
an imperfect Point Spread Function correction, but also from intrinsic
alignment of galaxies. The later effect has been observed for dark
matter halos in simulations, but it is still difficult to have a
reliable prediction of its amplitude. Nevertheless, it is
not believed to be higher than
a $10\%$ contribution for a lensing survey with a mean source
redshift at $z_s=1$.
In any case, intrinsic alignment
contamination can be removed completely by measuring the signal
correlation between distant redshift bins, instead of measuring
the full projected signal \cite{king02}.

\section{Cosmological Parameters Constraints and Conclusion}

As quoted earlier \cite{j97}, the cosmic shear signal depends primarily
on four parameters: the cosmological mean density $\Omega_m$, the mass power
spectrum normalization $\sigma_8$, the shape of the power spectrum $\Gamma$,
and the redshift of the sources $z_s$. Therefore, any measurement of the
statistics shown in Section 2 provide constraints on these parameters.
A big enough lensing survey \cite{huteg99}
will also provide constraints on many other parameters, but we do not discuss
this issue here.

\begin{figure}[t]
\centerline{
\psfig{figure=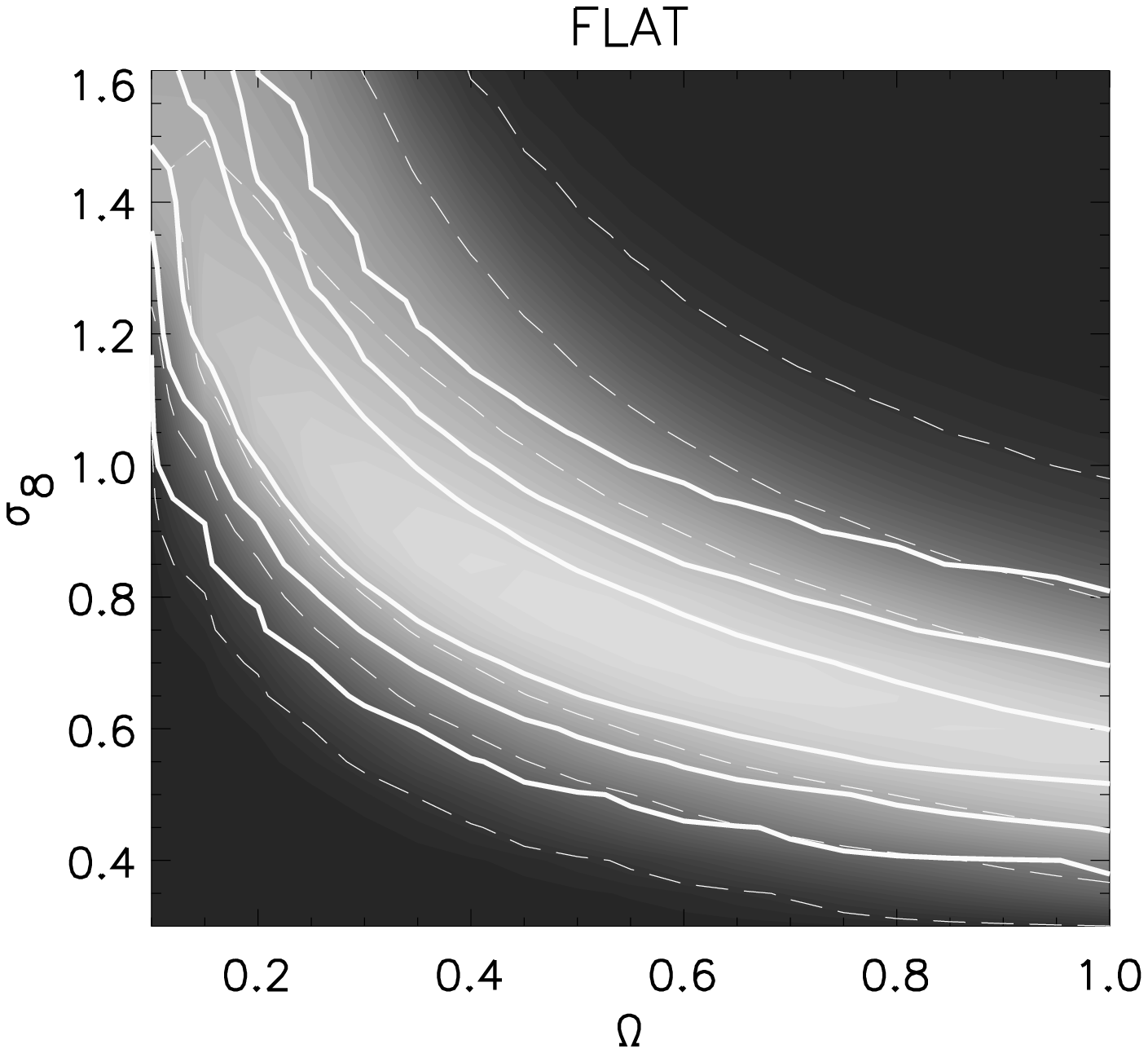,height=6.cm}
\psfig{figure=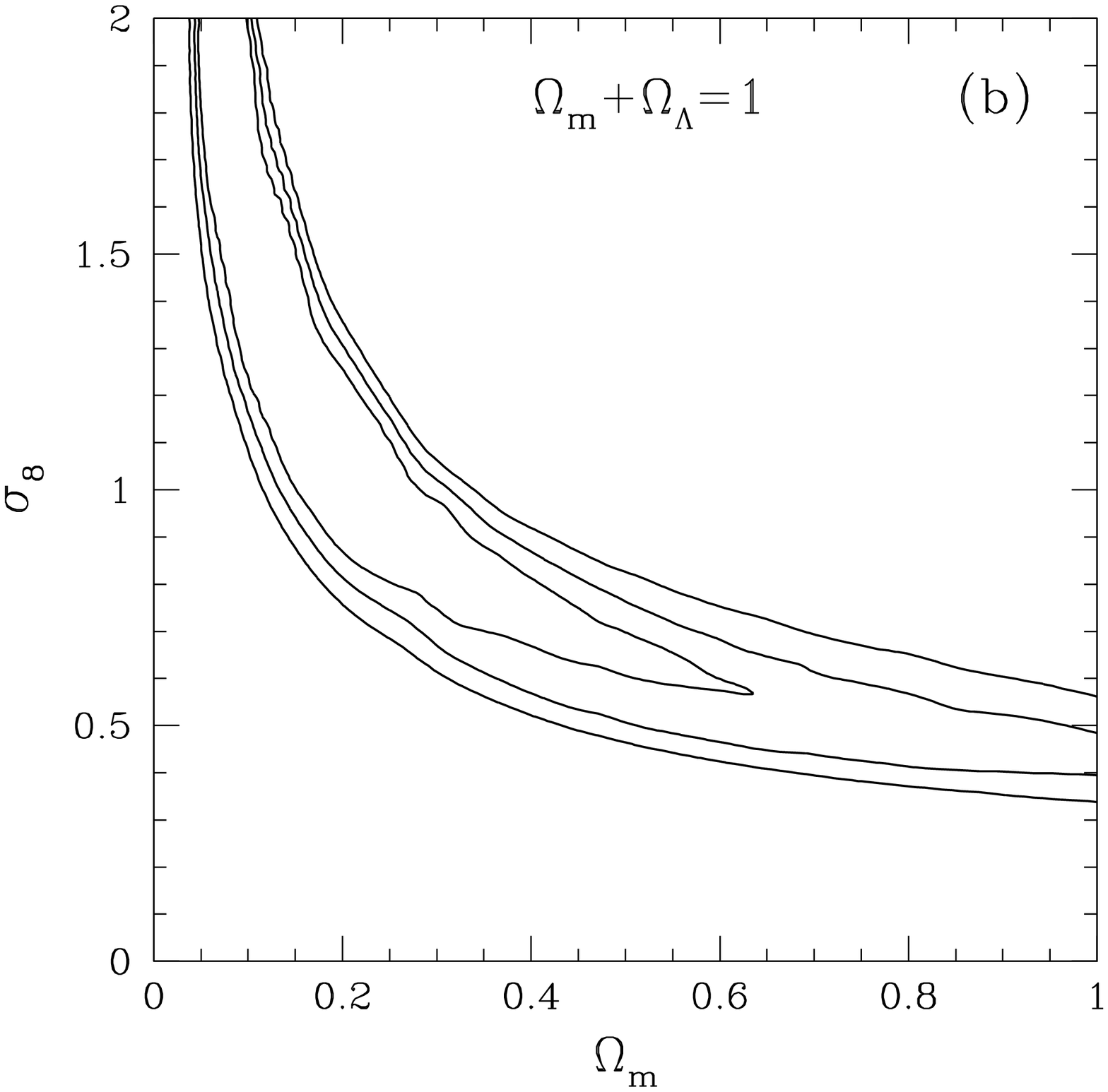,height=6.cm}
}
\caption[]{The solid lines on each plot show the 1, 2 and 3$\sigma$
contours of the VIRMOS and RCS survey, from the measurements shown in
Figure \ref{mapplot}. The contours have been marginalised over the source
redshift and the slope of the matter power spectrum as described elsewhere
\cite{vwal01,hoekstra02}.
\label{flat_omega_sigma8.ps}}
\end{figure}
Figure \ref{flat_omega_sigma8.ps} shows the joint
$\Omega_m$, $\sigma_8$ constraints
obtained from the measurements of Figure \ref{mapplot}. They are
obtained only when comparing the measured lensing signal to the non-linear
predictions. Indeed, the actual surveys are not yet big enough to probe the
linear scales accurately. The non-linear power can be computed numerically
\cite{smith02}, but its precision is still uncertain. Recent investigations
show that a $10 \%$ r.m.s. uncertainty is expected, which means that the
cosmological parameters cannot be known with better precision for the moment.
According to the Figures \ref{model.ps}, \ref{kernel.ps}, the transition scale
between the linear and non-linear regimes is around 1 degree. The consequence
is that
the quoted mass normalization $\sigma_8$ is sensitive to the validity
of the non-linear mapping at small scale. In this respect, \cite{jar02}
are less contaminated by this problem because they used the lensing signal
from $30'$ to $100'$ to constrain the mass normalization.

Table \ref{constraints} summarizes
the $\sigma_8$ measurements for all the lensing surveys published so far. For
simplicity it is given for $\Omega_m=0.3$. Despite the differences among the
surveys, it is worth to note that the results are all consistent within
$2.5\sigma$ between the most extreme cases, when poorly known parameters are
marginalised.

The residual $B$ mode
has been included in the analysis of the most recent cosmic shear surveys \cite{vwal02,
hoekstra02,ham02,jar02}, but we do not know yet how to deal with
it properly: some groups \cite{vwal02,hoekstra02,ham02} added the $B$ mode in
quadrature to the $E$ errors, taking into account the correlation between
various scales. The $B$ mode has been substracted first from the $E$ mode in
\cite{hoekstra02}, and not in \cite{vwal02}. This might probably result in a slight
bias for high $\sigma_8$ values in \cite{vwal02}. Unfortunately we have no garantee
that the $B$ substraction is the right correction method.
Recently \cite{jar02} marginalised the probabilities
over $E-B$ to $E+B$ taken as the signal, which is more likely to include
the 'true' $B$ mode correction one has to apply.

The $B$ mode correction
and the non-linear power spectrum predictions are, today, the major limitations
of cosmic shear surveys. They prevent the measurements of cosmological
parameters below $10\%$, even if today's lensing surveys could do
better in principle. It is therefore very difficult to establish a
discrepancy among the cosmic shear results today. There is probably something
to learn about basic PSF
correction with the existing data, since the different groups get different
$B$ mode amplitude and scale dependence. This is still widely unexplored.

The positive aspect is that cosmic shear is now established as a powerfull
tool to probe the dark matter and the cosmological parameters, and that
it works, which was not the case only 2 years ago. There is a lot to
expect from forthcoming surveys, with the first results expected in less than
a year from now.


%
\acknowledgements{We thank H. Hoekstra, U.-L. Pen, P. Schneider for
useful discussions. We thank Anthony Lewis for the use of the CAMB software.
   This work was supported by the TMR Network
``Gravitational
 Lensing: New Constraints on
Cosmology and the Distribution of Dark Matter'' of the EC under contract
No. ERBFMRX-CT97-0172.
}

\vfill
\begin{landscape}
\begin{table*}[t]
\begin{center}
{\small
\caption{Constraints on the power spectrum normalisation "$\sigma_8$" for
$\Omega_m=0.3$ for a flat Universe, obtained from a given "statistic".
"CosVar" tells us whether or not the cosmic variance
has been included, "E/B" tells us whether or not a mode decomposition has been used
in the likelihood analysis. Note that \cite{vwal01} and \cite{brown02} measured a
small B-mode, which they didn't use in the parameter estimation. $z_s$ and
$\Gamma$ are the priors used for the different
surveys identified with "ID". Note also the cosmic shear results obtained by
\cite{kais00,hammerle01}, which
are not in the table here because they reported a shear detection, not a $\sigma_8$
measurement.
\label{constraints}
}
\label{tabcs}
\bigskip
\begin{tabular}{lcccccccc}\hline
\\
ID & $\sigma_8 $ & Statistic & Field & $m_{\rm lim}$& CosVar & E/B & $z_s$ & $\Gamma$ \\
\\
\hline
\\
\cite{mao01} Maoli & $1.03\pm 0.05$ & $\langle\gamma^2\rangle$ & VLT+CTIO+ & - & no & no & - & 0.21 \\
et al. 01 &  &  & WHT+CFHT & & & & & \\
\\
\hline
\\
\cite{vwal01} LVW& $0.88\pm 0.11$ & $\langle\gamma^2\rangle$, $\xi(r)$  & CFHT & I=24 & no & no & 1.1 & 0.21 \\
et al. 01 &  & $\langle M_{\rm ap}^2\rangle$ & 8 sq.deg. &  & & (yes) &  &  \\
 \\
\hline
\\
\cite{rhodes01} Rhodes & $0.91^{+0.25}_{-0.29}$ & $\xi(r)$ & HST & I=26 & yes & no & 0.9-1.1 & 0.25 \\
et al. 01 &  &  & 0.05 sq.deg. & & & &  & \\
 &  &  &  & & & & & \\
 \\
\hline
\\
\cite{hoek01a} Hoekstra& $0.81\pm 0.08$ & $\langle\gamma^2\rangle$ & CFHT+CTIO & R=24 & yes & no & 0.55 & 0.21 \\
et al. 01 &  & & 24 sq.deg. &  & &  &  &  \\
\\
\hline
\\
\cite{bacon02} Bacon& $0.97\pm 0.13$ & $\xi(r)$ & Keck+WHT & R=25 & yes & no & 0.7-0.9 & 0.21 \\
et al. 02 &  &  & 1.6 sq.deg. & & & & & \\
\\
\hline
\\
\cite{ref02} Refregier & $0.94\pm 0.17$ & $\langle\gamma^2\rangle$ & HST & I=23.5 & yes & no & 0.8-1.0 & 0.21 \\
et al. 02 &  &  & 0.36 sq.deg. & & & &  & \\
\\
\hline
\\
\cite{vwal02} LVW& $0.94\pm 0.12$ & $\langle M_{\rm ap}^2\rangle$ & CFHT & I=24 & yes & yes & 0.78-1.08 & 0.1-0.4 \\
et al. 02 &  &  & 12 sq.deg. &  & &  &   &   \\
\\
\hline
\\
\cite{hoekstra02} Hoekstra& $0.91^{+0.05}_{-0.12}$ & $\langle\gamma^2\rangle$, $\xi(r)$ & CFHT+CTIO & R=24 & yes & yes & 0.54-0.66 & 0.05-0.5 \\
et al. 02 &  & $\langle M_{\rm ap}^2\rangle $ & 53 sq.deg. &  & &  &  &  \\
 \\
\hline
\\
\cite{brown02} Brown&  $0.74\pm 0.09$ & $\langle\gamma^2\rangle$, $\xi(r)$  & ESO & R=25.5 & yes & no & 0.8-0.9 & - \\
et al. 02 &  &  & 1.25 sq.deg. & & & (yes) &  &  \\
\\
\hline
\\
\cite{ham02} Hamana & $(2\sigma) 0.69^{+0.35}_{-0.25}$ & $\langle M_{\rm ap}^2\rangle$, $\xi(r)$  & Subaru & R=26 & yes & yes & 0.8-1.4 & 0.1-0.4 \\
et al. 02 &  &  & 2.1 sq.deg. & & &  &   & \\
 \\
\hline
\\
\cite{jar02} Jarvis & $(2\sigma) 0.71^{+0.12}_{-0.16}$ & $\langle\gamma^2\rangle$, $\xi(r)$  & CTIO & R=23 & yes & yes & 0.66 & 0.15-0.5 \\
et al. 02 &  & $\langle M_{\rm ap}^2\rangle$ & 75 sq.deg. & & &  &  &  \\
 \\
\hline
\end{tabular}
}
\end{center}
\end{table*}
\end{landscape}

\end{document}